\documentclass[a4paper]{jpconf}
\usepackage{graphicx}
\begin{document}
\title{
Unconventional magnets in external magnetic fields} 

\author{Roderich Moessner}

\address{Max-Planck-Institut f\"ur the Physik of komplexer Systeme, Dresden,
  GERMANY} 

\ead{moessner@pks.mpg.de}

\begin{abstract}
This short review surveys phenomena observed when a magnetic field is applied
to a system of localised spins on a lattice. Its focus is on frustrated magnets
in dimension $d \geq 2$. The interplay of field and entropy is illustrated in
the context of their unusual magnetocaloric properties, where field-tuned
degeneracies assert themselves. Magnetisation plateaux can reveal the physics of
fluctuations, with unusual excitations (such as local modes, extended
string defects or monopoles) involved in plateau termination. Field-tuning
lattice geometry is the final topic, where mechanisms for dimensional reduction
and conversion between different lattice types are discussed.
\end{abstract}

\section{Introduction}
\label{sec:Intro}

The attraction of studying magnets arises from a combination of three
factors. Firstly, magnetic models often provide the simplest setting for the
study of complex cooperative phenomena. This has for instance permitted the
development of detailed theories of ordering phenomena linking microscopics
with thermodynamics and even out-of-equilibrium behaviour. Secondly, a
seemingly unlimited range of experimental compounds \cite{schiframir} in fact
realises a wide range of these phenomena, the study of which constantly not only
throws up new discoveries and but also tests existing theoretical
understanding. 

Thirdly, a closely connected factor is the availability of a wide variety of
probes for studying magnets experimentally. Too numerous even to list here
exhaustively, it is nonetheless worth emphasizing how varied their nature is,
ranging from local ones (nuclear magnetic resonance or muon spin rotation) to
the usual thermodynamic ones. Worthy of particular mention are neutrons, which
provide information on essentially the full dynamical structure factor in a
straightforward way, their magnetic coupling being weak enough not to perturb
the system substantially under usual circumstances. The advent of high
resolution X-rays capable of providing complementary information about
microscopic magnetic structure is one of the current exciting experimental
developments. 

Magnetic fields provide another very convenient handle for studying and
manipulating magnets in a controlled and moderately non-invasive way. Often,
but by no means always, the energy scales are such that readily available
laboratory fields, of the order of 10 Tesla, are comparable to thermal and
exchange energies. Dedicated facilities can provide much higher magnetic
fields. In particular, pulsed fields routinely achieve strengths well above 50
T. However, in this regime, other parameters such as width and shape of the
field pulse play an important role; for example, for very 
short pulses, equilibration is not always straightforward. Besides, just as
impressive as data from successful high-field experiments can be, the aftermath
of a failed one is also spectacular, but certainly such an experiment no larger qualifies as non-invasive \cite{NHMFL}. 

The behaviour of magnets in an applied magnetic field is the subject of this
brief review. It aims to expose conceptually important phenomena in as simple a
setting as possible, using well-studied model systems for illustrative
purposes. Emphasis is placed on magnets in dimension of two or higher, in
particular magnets the low-temperature behaviour of which in zero field is
already sufficiently exotic. This includes in particular classes of frustrated
magnets \cite{pub:frustoday} where a cooperative paramagnetic or spin liquid
regime extends down to very low temperatures \cite{CJP}.

We set up the problem in the following section. Section \ref{sec:Magnetisation}
is devoted to magnetisation plateaux, including how they are stabilised, and
how they terminate. In Section \ref{sec:TunLattice}, we discuss the influence a
magnetic field can have on the effective geometry of the magnetic system,
including specific examples of dimensional reduction, and the generation of
unusual defects.

\section{The Hamiltonian}
\label{sec:Hamil}

Let us consider spins $\vec{S}$, with $i = 1 ... N$, residing on a regular lattice. We
will consider both isotropic (Heisenberg) interactions as well as uniaxial
(Ising) ones: 
\begin{eqnarray}
\mathcal{H}_\mathcal{H} & = & J \sum_{(ij)} \vec{S}_i\cdot \vec{S}_j\nonumber \\
\mathcal{H}_I & = & J \sum_{(ij)} S^z_i S^z_j~~~,
\label{eq:1}
\end{eqnarray}
where the sum runs over all bonds of the lattice.

In addition, there is the magnetic field term, which can in its simplest
incarnation be written as 
\begin{equation}
\mathcal{H}_B = -\sum^N_{i=1} \vec{\mu}_i \cdot \vec{B} = -\sum^N_{i=1} \vec{h}
\cdot \vec{S}_i 
\label{eq:2}
\end{equation}
where the magnetic moment $\vec{\mu} = g \mu_B \vec{S}$ related to the spin via
the product of $g$-factor (assumed to be isotropic), and the Bohr magneton,
$\mu_B$. In the following, we will use the reduced field $\vec{h} = g \mu_B
\vec{B}$ for notational simplicity. 

In the case of Heisenberg interactions, the field direction defines a
privileged axis and thus reduces the symmetry of the Hamiltonian from SU(2) to
U(1). The situation for Ising interactions can be much more complex. In particular,
the local $z$-directions need not be identical; as we will discuss in detail
below, this can lead to a wide variety of interesting phenomena when a magnetic
field is applied. 

\section{Magnetisation plateaux, and what happens around the edges}
\label{sec:Magnetisation}

A linear response of the magnetisation, $\vec{m}$, to an applied field allows
the definition of a susceptibility, $\chi = \frac{dm}{dh}$. For classical
magnets at low temperature, one might expect the linear regime to be very
broad. To see this, let us minimise $\mathcal{H}_\mathcal{H} + \mathcal{H}_B$
to find ground states in an applied field. By denoting $\vec{L}_{ij} =
\vec{S}_i + \vec{S}_j$, and for a lattice of uniform coordination $z$, we can
complete the square to write 
\begin{equation}
\mathcal{H}_\mathcal{H} + \mathcal{H}_B = \frac{J}{2} \sum_{ij} \left[ \left(
\vec{L}_{ij} - \frac{\vec{h}}{zJ} \right)^2 - \left( \frac{h}{zJ} \right)^2
\right] ~~~. 
\label{eq:3}
\end{equation}

From this, it would follow that $\vec{m} = \frac{1}{2} \vec{L}_{ij} = \chi_0
\vec{h}$, with $\chi_0^{-1} = 2zJ$ for all strengths of the magnetic field. 

Nonetheless, it is frequently observed that the magnetisation curve as a
function of field is not at all uniformly linear. The reasons for this can be
very varied -- the usual suspects are thermal or quantum fluctuation, or
additional terms in the Hamiltonian omitted in Sec. (\ref{sec:Hamil}).

\subsection{The saturated plateau}
\label{sec:SatPlat}

Indeed, the simplest instance of a magnetisation plateau occurs when the field
is strong enough to saturate the magnetisation: $\vec{m} = \vec{S}$. Beyond
this, it is clearly impossible to increase the magnetisation (at any rate
without the involvement of some high-energy process such as rearranging the
crystal-field scheme). Simple though it is, this plateau has some very useful
and remarkable properties.

For quantum spins, the saturation plateau is useful as it reveals properties of
the Hamiltonian in a moderately direct way. This happens because the ground
state is a simple, completely unentangled state with all spins in the state 
$S^z = S$: 
\begin{equation}
| {\rm sat} \rangle = \otimes^N_{i=1} | S^z_i = S \rangle~~~. 
\label{eq:4}
\end{equation}
This is in contrast to the N\'eel state, for example, which for finite spin $S$
is always dressed with quantum fluctuations. Starting from this state, it is
then easy to construct excitations. For instance, a spin wave with momentum
$\vec{k}$ will have an energy above the field-induced gap given by the Fourier
transform of the exchange integral, $J(\vec{k})$:
\begin{equation}
\mid \vec{k} \rangle = \frac{1}{\sqrt{N}} \sum e^{i \vec{k} \vec{r}_i} S^-_i
\mid {\rm sat} \rangle \Longrightarrow \langle \vec{k} \mid \mathcal{H} \mid
\vec{k} \rangle \sim J (\vec{k})+\mathrm{const}~~~.  
\label{eq:5}
\end{equation}
Here, $S^-$ is the spin lowering operator. 

By measuring the spin wave excitation spectrum in inelastic neutron scattering
across the Brillouin zone, it is thus possible to extract reliable values of
the exchange constants. However, this method is still not entirely
model-independent as different choices of terms in $\mathcal{H}$ can lead to
similar shapes of $J(\vec{k})$, so that one still needs to use educated
guesses, e.g. based on quantum chemical and exchange pathway considerations, to
set up a model Hamiltonian.  

One example of this is provided by a study of Cs$_2$CuCl$_4$ \cite{coldea}, a
stacked anisotropic triangular lattice magnet exhibiting an unusual excitation
spectrum in that there appear to be broad neutron scattering continua rather
than only well-defined magnon peaks (for recent theoretical work on this, see
\cite{STARBAL1} and references therein). Besides
obtaining values for the leading nearest-neighbor exchange, the shape of the
dispersion relation together with symmetry considerations led to the conclusion
that -- besides next-nearest and interlayer couplings -- there must also be a
Dzyaloshinskii-Moriya interaction, which arises from spin-orbit coupling.

In passing we note a different setting where this idea has been developed
further. In the case of ferromagnets (which of course also have a
fluctuation-free fully polarised ground state), one can make a fair amount of
progress in solving the two-excitation problem, because the single spin-flip
ones are exactly known. This information is used to infer properties
of the ordered state, which can be exotic if the bound state of two magnons
condenses before the single magnons do \cite{shannon1}.

How does such a plateau terminate as the external field is reduced? The
simplest scenario for this to occur is via the condensation of the softest
excitation. In the case of a square lattice antiferromagnet, this
would be a magnon at wavevector $(\pi, \pi)$, and its condensation leads to the
symmetry breaking pattern characteristic of a N\'eel date. As the condensation
of this bosonic mode breaks the U(1) symmetry of rotations about the axis
defined by the field, one can construct a formal analogy between this
transition and Bose-Einstein condensation. This is explained in detail in the
review \cite{TcherSS}. 

\begin{figure}[h]
\includegraphics[width=18pc]{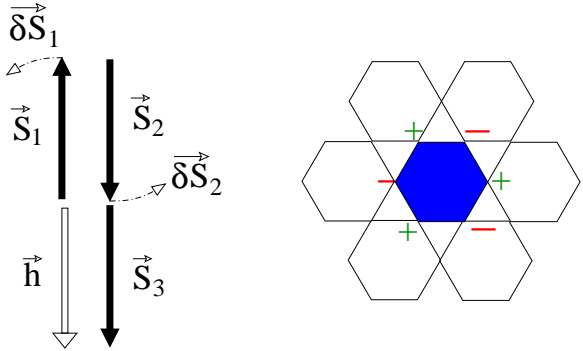}\hspace{2pc}%
\begin{minipage}[b]{14pc}\caption{\label{locmag} Special excitations on the
	kagome lattice. Left: Soft fluctuation around a collinear field-induced
	state. Right: Pattern of a low-lying local magnon near saturation. A magnon
    on one hexagon penalises the presence of further magnons on neighbouring
    hexagons.} 
\end{minipage}
\end{figure}

In a frustrated magnet, there is generically no single softest mode. Rather,
there can be an entire band of minimal energy spinwaves, stretching through the
full Brillouin zone. This physics is most easily illustrated in two dimensions
for the case of the kagome lattice \cite{Schulenburg}, where the flat bands are
due to local magnon modes around a hexagon.

Such a local magnon mode is displayed in Fig. (\ref{locmag}). This mode is
local because like in an Aharonov-Bohm cage \cite{ABcage} the staggered
magnetisation around the hexagon gives rise to a cancellation in the exchange
field on the triangular sites capping the hexagon. To leading order, the number
of independent such modes equals the number of unit cells of the lattice: there
is no way of choosing  subsets $A$, $B$  of the full lattice
$\Lambda$ made
up of the sites contained in disjoint sets of hexagons such that $A = B$ and
with $A \cup B$ a proper subset of $\Lambda$. Only when $A \cup B = \Lambda$ does
this become possible, which implies $N_{\rm modes} = N_{\rm unit cells} - 1$
\cite{SchmiMoRi,ZhTs1}. However, there are two further modes
winding around the system (for periodic boundary conditions), which gives
$N_{\rm modes} = N_{\rm unit cells} + 1$, in keeping with the presence of a
flat band and a gapless dispersive band \cite{bergmann}.

At the saturation field, the Zeeman energy cost of reducing $S^z$ is exactly
offset by the exchange energy gain of the staggered transverse magnetic
moment. As long as the magnon modes do not overlap, their energies are simply
additive. By contrast, overlapping modes incur an additional energy cost
\cite{Schulenburg}, so that there is a degeneracy given by the number of ways
of putting non-overlapping hexagons onto the kagome lattice. The entropy of the
resulting hard-hexagon model is known exactly to be $S \approx 0.1111$
\cite{Baxter}. This entropy can lead to impressive magnetocaloric effects,
which we will discuss an example of a simpler setting below. Unusual
magnetocaloric effects due to such anomalous densities of states in frustrated
magnets are quite common overall -- see Reference \cite{ZhitoMC} and citations
thereof. 

Let us next consider what happens when the field is lowered further. As the
field is turned through the saturation field, the energy of the local magnons
goes negative, and they thus obtain their maximal density. As magnons 'repel'
-- the magnetisation therefore jumps to $S - N_{\rm max}/N_{\rm
	sites}$, where $N_{\rm max}$ is the maximal number of non-overlapping
hexagons, which equals $N_{\rm sites}/9$.  
This magnetisation jump is independent of $S$, and therfore its relative size
vanishes in the semiclassical limit, but it is very significant for small $S =
\frac{1}{2}$! 

\subsection{Intermediate plateaux}
\label{sec:IntPlat}

The genesis of the saturated plateau is considerably more simple than that of
fractional plateaux at, say, 1/2 or 1 /3 of the saturated magnetisation. These
constitute a subject in their own right, and as their origin is very varied, we
can only give some illustrative examples here.

We first discuss fluctuation-induced collinear plateaux. By
fluctuation-induced, we mean that, whereas energetically a linear
$\vec{m}(\vec{h})$ curve would be favourable in a classical magnet, thermal or
quantum fluctuations instead favour collinear states. However, collinear states
are not always available given the fixed length of the individual spins. For
this reason, the  values of $\vec{m}_p$ for which collinear states {\em do}
exist are stabilised by fluctuations over a finite range of fields 
around $\vec{h}_p=\vec{m}_p/\chi_0$.  

\subsubsection{Thermal plateaux}
\label{sec:ThermPlat}

For thermal fluctuations, this represents a typical instance of classical order
by disorder \cite{CJP}. This happens, for instance, in the kagome magnet
\cite{ZhitoKP}, the zero-field behaviour of which presents a celebrated example
of thermal order by disorder in magnetism \cite{ChaHold,RCC,HKB,HuseRut}. In a
nutshell, the frequencies for some 
fluctuations around collinear states vanish in the harmonic approximation (see
Fig. \ref{locmag}). These `soft modes' in fact typically have a quartic energy, $E
\propto \delta^4$, so that by equipartition $\langle \delta^2 \rangle \propto
\sqrt{T}$, which is much larger that the $\langle \delta^2 \rangle \propto T$
for quadratic modes at low T. 

\begin{figure}[h]
\includegraphics[width=14pc]{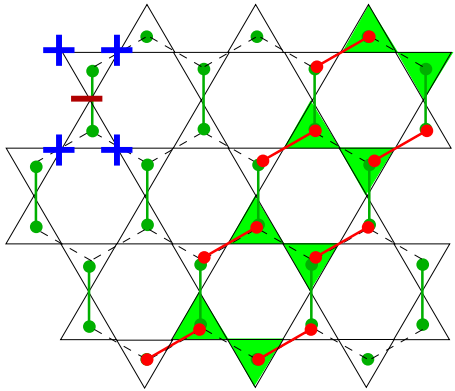}\hspace{2pc}%
\begin{minipage}[b]{14pc}\caption{\label{fig:string}Mapping from (down) spins
	on the kagome lattice to (occupied) dimers on the honeycomb. A dimer
    rearrangement leaving the zero-point energy invariant is also shown. The
    states shown are favoured by quantum fluctuations in an applied field.}
\end{minipage}
\end{figure}

This means that the fluctuations out of collinear states have a much larger
amplitude than for non-collinear ones, and hence they lead to a lower free
energy for the former. As this free energy gain is of entropic origin, it
vanishes in the limit of low temperature, and the width of the plateau -- where
the entropic gain outweighs the energetic cost of departing from $M \propto H$
-- vanishes with T. As this plateau corresponds to two spin pointing along the
field and one in the opposite direction, it occurs at $m_p = m_{\rm sat}/3$.

\subsubsection{Semiclassical plateaux}
\label{subsec:SemPlat}

Quantum fluctuations around classical states provide another mechanism of
stabilising magnetisation plateaux \cite{ChubuP}. This follows from the fact
that, semiclassically, one can use the commutation relationship
$[\hat{S}_x,\hat{S}_y] = i \hbar \hat{S}_z$ to obtain a perturbative treatment
of transverse fluctuations. With the classical orientation defining the local
$z$-direction, $\hat{S}^z \equiv S$, the relative transverse fluctuations are
parametrically small in $1/S$: 
\begin{equation}
\left[ \frac{\hat{S}_x}{S}, \frac{\hat{S}_y}{S} \right] = \frac{i \hbar}{S}~~~.
\label{eq:6}
\end{equation}

Using 1/S as a small parameter yields an expansion of the total energy,
schematically given by 
$E = E_{\rm classical} +  E_{\rm semiclassical} +$ higher
order terms, where $E_{\rm classical}$ is of $O(S^2)$, and $E_{\rm
  semiclassical} \sim \sum_i \hbar \omega_i$, of $O(S)$, encodes the zero-point
energy of the excitations around the classical state.  

Unlike the thermal fluctuations discussed above, where the soft modes gave rise
to the stability of the plateaux, here, {\it all} modes contribute. This makes
a controlled analytical treatment much harder. A long time ago, Henley has
argued that the net effect of fluctuations is to induce an effective
biquadratic exchange,
\begin{equation}
{\cal H}_{biq}\sim -\frac{J}{S}\sum_{(ij)}\left(
\vec{S}_i\cdot\vec{S}_j
\right)^2
\end{equation}
 favouring collinearity. For the latest analysis of this
issue, see \cite{LarHen}. This has been studied explicity for the kagome
lattice in a field, where a monotonic dependence of zero-point energy on
collinearity was found for a set of three sublattice states, but of a more
complicated form, than a simple biquadratic exchange \cite{HasMoe}.

However, this is not the end of the story: there are in fact many different
collinear ``2-up 1-down'' states with; their number can  be determined
exactly, because these states map onto a soluble problem in classical
statistical mechanics: the hexagonal lattice dimer model (Fig.~\ref{fig:string}). 
The hexagonal lattice
is the lattice defined by the centers of the triangles of the kagome lattice;
the sites of the latter map onto the bonds of the former. Colouring in the
(hexagonal lattice) bonds corresponding to a down spin yield a hardcore dimer
covering. 

Generalising an approach due to Henley \cite{Henley} to include
magnetic fields, one needs to evaluate the zero-point energy given by
\cite{HasMoe}
\begin{equation}
E_{\rm semicalssical} = Tr \sqrt{\left( \frac{h^2}{4} + 3 \right)
  \delta_{\alpha \beta} + hS_{\alpha \beta} + \left( S_{\alpha \gamma}
  S_{\gamma \beta} - 3 \delta_{\alpha \beta} \right)}~~~,
\label{eq:7}
\end{equation}
where $S_{\alpha \beta}$ is the spin connecting shared by triangles $\alpha,
\beta$ of the kagome lattice.

One finds that a uniform ``$q = 0$'' state, where all dimers are parallel,
  minimises $E_{\rm semicalssical}$ (Fig.~\ref{fig:string}). This is somewhat
  surprising as, up until this result \cite{kagfrench}, a ``$\sqrt{3} \times
  \sqrt{3}$'' state with an enlarged unit cell had always been the
  fluctuation-favoured state in a wide variety of settings. However, related
  models of quantum fluctuations -- such as one for easy-axis spins
  \cite{KedarEA} -- have yielded a similar ordering pattern since.

This is not yet the end of the story. There in fact remains a degeneracy --
exponential in the linear size of the system -- involving configurations where
an even number of columns of dimers have been shifted
(Fig. \ref{fig:string}). This results from Henley's gaugelike symmetry
\cite{Henley}, and is discussed in detail in \cite{HasMoe}, or, in an analogous
zero-field setting \cite{otchslwc}.
 
\subsubsection{Magnetoelastic plateaux}
\label{subsec:MagElPlat}

A conceptually attractive route to a magnetisation plateau has recently been
prompted by experiments on the spinel CdCr$_2$O$_4$ \cite{TakPlat,ShaPlat},
where the isotropic $S = 3/2$ Cr ions reside on a pyrochlore lattice
(Fig. \ref{fig:lattune}). Here, the plateau is stabilised via coupling of the spins to
elastic lattice degrees of freedom. In the simplest model, these are captured
by two additional terms, $\mathcal{H}_e$ encoding the harmonic restoring force
for deviations $x_i$ of the ions from their equilibrium position, and
$\mathcal{H}_{me}$, which models the distance dependence of the magnetic
exchange via a simple Taylor expansion:
\begin{equation}
\mathcal{H}_e = \frac{k}{2} \sum_{(ij)} \left( x_i - x_j \right)^2
~~;~~~\mathcal{H}_{me} = J' \sum_{(ij)} \left( x_i - x_j \right) \vec{S}_i
\vec{S}_j~~~. 
\label{eq:8}
\end{equation}
The phonons can then be straightforwardly integrated out (see e.g. \cite{LarHen}), 
and
one obtains again a biquadratic effective exchange, symbolically written as:
\begin{equation}
\mathcal{H}^{me}_{\rm eff} \sim -\frac{J'^2}{k} \left( \vec{S}_i
\vec{S}_j \right)^2~~~. 
\label{eq:bieff}
\end{equation}
The further analysis of course does not depend on the origin of this
biquadratic exchange. In closing, we note that frustrated magnets are
inherently susceptible to magnetodistortive transitions as their many
ground-states include non-uniform patterns of the bond energy $\vec{S}_i \cdot
\vec{S}_j$, and $\mathcal{H}^{me}_{\rm eff}$ (\ref{eq:bieff}) thus has a
nontrivial action in the space of degenerate states to first order in its
prefactor \cite{YaUe,TchME}.

\subsubsection{Commensurability plateaux}
\label{subsec:CommPlat}

There is a long-standing strand of activity centered around the
``Shastry-Sutherland'' \cite{ShasSuther} 
compound SrCu$_2$(BO$_3$)$_2$, where various
intermediate plateaux have been, and are being, discovered
experimentally. Their theoretical
study is being pursued under the heading of supersolid magnetic phases. This
field is too broad, and too much in flux, to be given justice here, and we
refer the interested reader to an upcoming review \cite{Takigawa}.

The basic idea again is to treat excitations as weakly interacting 
Bosons \cite{TcherSS}. At commensurate densities $\rho = p/q$ (where $p$, $q$
are small integers), a Mott transition for these Bosons can occur, the
incompressibility of which manifests itself as a magnetisation plateau.

Incompressible states are of course not exclusive to Bosons -- indeed, Fermions
in an external field exhibit the fractional quantum Hall effect, itself one of
the most prominent incompressible states. Based on a loose analogy to this
physics, Misguich et al. \cite{MJG} have developed a Fermionic theory of the
magnetisation plateau. The status of the connection between Fermionic and
Bosonic theories is at present unclear. 

This concludes our survey of Heisenberg magnetisation plateaux. The next
section, which deals with the interplay of magnetic field and lattice geometry,
we will encounter another plateau, which goes along with a reduction of the
effective lattice dimensionality.

\section{Tuning lattice geometry via a magnetic field}
\label{sec:TunLattice}

To demonstrate the degree of control magnetic fields provide, this section
discusses how they can be used even to modify the effective lattice
geometry. In the following, we show how the lattice type can be tuned
(pyrochlore to face-centered cubic), how its dimensionality can be modified
(dimensional reduction from $d = 3$ to $d = 2$ or $d = 1$), and how the nature
of the low-energy degrees of freedom changes along the way; we will
e.g. encounter extended string defects and spins fractionalising into
monomers/monopoles.  

\begin{figure}[h]
\includegraphics[width=18pc]{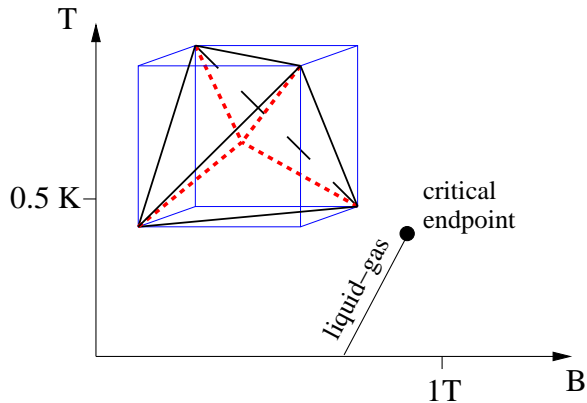}\hspace{2pc}%
\begin{minipage}[b]{14pc}\caption{\label{fig:liquidgas}Liquid-gas phase diagram
	of spin ice  with long-ranged dipolar interactions in a magnetic field. For
	nearest neighbour interactions only, the liquid-gas phase transition gets
	replaced by a crossover. Inset: non-collinear Ising axes (dashed lines) of
	spins on a tetrahedron.} 
\end{minipage}
\end{figure}

The mechanism whereby this is achieved involves coupling the field selectively
to certain subsets of spins. The central ingredient to our example is a
classical Ising model in which the Ising axes are non-collinear
\cite{RMR}. This can be realised for four spins arranged on the corners of a
tetrahedron, which has four such non-collinear axes (see inset of 
Fig. \ref{fig:liquidgas}). These are the cubic [111]
axes, $\hat{d}_\kappa$ ($\kappa$ = 1...4), which all have a mutual angle,
$\theta$, with $\cos \theta = \hat{d}_k \cdot \hat{d}_k' = -1/3$.   

\begin{figure}[h]
\includegraphics[width=18pc]{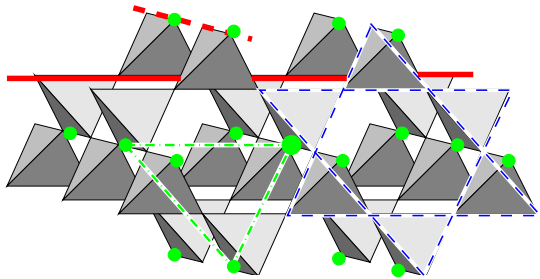}\hspace{2pc}%
\begin{minipage}[b]{14pc}\caption{\label{fig:lattune}The pyrochlore
	lattice. One sublattice (green dots) forms a face-centred cubic
    lattice. Orthogonal chains are shown as red lines (solid and dashed). One
    kagome star in a \{111\} plane is shown (dashed blue), along with a
    triangle belonging to the alternating triangular plane (dashed green).}
\end{minipage}
\end{figure}

This non-collinearity has as its first striking consequence that a ferromagnet
is frustrated, and an antiferromagnet is not. Here, by (anti)ferromagnet, one
means a positive (negative) Curie temperature, $\Theta \propto z J$, obtained
from the inverse susceptibility $\chi^{-1} \propto T - \Theta$ at high
temperature $T \gg |\Theta|$. Introducing a variable $\sigma_i = \pm 1$ which
encodes if a spin points into or out of the tetrahedron along its Ising axis,
one finds 
\begin{equation}
\sum_{\langle ij \rangle} J \vec{S}_i \cdot\vec{S}_j = J \sum_{\langle ij \rangle}
\left( \sigma_i \hat{d}_{\kappa(i)} \right)\cdot \left( \sigma_j
  \hat{d}_{\kappa(j)} \right) = -\frac{J}{3} \sum_{\langle ij \rangle} \sigma_i
\sigma_j ~~~. 
\label{eq:10}
\end{equation}

Hence, the low-temperature is described by an Ising model with $J_{\rm eff} =
-J/3$. for a ferromagnetic $J < 0$, one thus obtains an antiferromagnetic
$J_{\rm eff}$. For the single tetrahedron, $-\frac{J}{3} \sum_{\langle ij
  \rangle} \sigma_i \sigma_j = -\frac{J}{6} \left( \sum^4_{i=1} \sigma_i
\right)^2 + \frac{2J}{3}$ there are thus $\left( {4 \atop 2} \right) = 6$
ground states, those with two spins pointing in, and two out. 

Now, consider applying a magnetic field along one of these axes, so that its
projection onto the spin on the corresponding `sublattice' ($i = 1$, say) is
three times that onto the others. A magnetic field which is much smaller than
$J$ will lift the degeneracy of the six ground states. Having fixed the
direction of one spin with the field ($\sigma = 1$, say), exactly one of the
three remaining spins must have the same value $\sigma = 1$, and the ground
state degeneracy is thus halved. 

\begin{figure}[h]
\includegraphics[width=18pc]{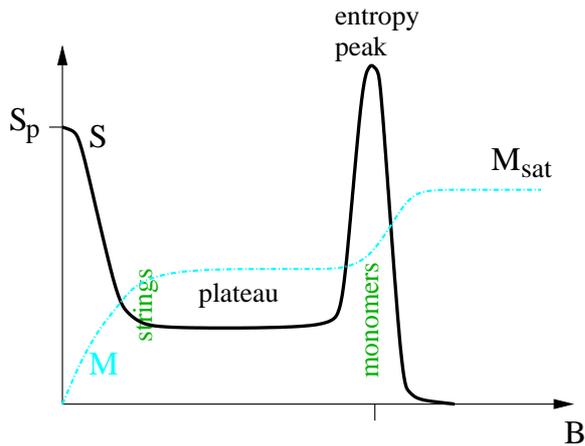}\hspace{2pc}%
\begin{minipage}[b]{14pc}\caption{\label{SIthermo}Schematic behaviour of
    magnetisation and entropy for spin ice in a [111] field. The central
    ``kagome ice'' plateau is terminated by string and monomer defects at low
    and high fields, respectively.}
\end{minipage}
\end{figure}

Next, let us combe these tetrahedra to form a lattice. A particularly
productive way of doing this is to arrange the tetrahedra so that they share
corners (Fig. \ref{fig:lattune}), with their centers forming a diamond lattice. The
resulting pyrochlore lattice decomposes into the pinned sublattice, which lie
on a face-centred cubic lattice. This can be thought of as stacks of triangular
layers, alternating with kagome layers hosting the  spins on the other three
sublattices. If interactions exist only between nearest
neighbours, these layers are decoupled thanks to the inert triangular
layers--dimensional reduction from $d=3$ pyrochlore to $d=2$ kagome has
occurred. This field-tuned dimensionality is also evident in the magnetic 
correlations \cite{SI1,fenbram}

With each triangle harbouring exactly one spin with $\sigma = +1$, we retrieve
the mapping to a dimer model on the hexagonal lattice described above in the
context of semiclassical plateaux. Hence, the plateau exhibits a finite entropy
{\em per spin of the pyrochlore lattice} of $S \approx 0.08 k_B$.

This entropy is different from that of the zero-field case, which has an
entropy per spin of $S_p \approx \frac{1}{2} \log \frac{3}{2}$. This ``Pauling
entropy'' was derived by Linus Pauling in the context of ice, and indeed the
Ising pyrochlore antiferromagnet is known as spin ice: tetrahedra with two
spins pointing in and two pointing out are said to obey the ice rule. This
model was first discussed by Anderson in 1956 \cite{SIA}, with the
corresponding compounds discovered in 1997 \cite{SIdis}, and the Pauling
entropy measured shortly thereafter \cite{SIent}. There has been a sustained
research activity ever since, which we will have to review on a separate
occasion. 

How is three-dimensionality re-established when the field strength is lowered
at low $T$? This happens via string defects, which are themselves one
dimensional. Their extended nature follows from a simple consideration: the
tetrahedra pointing upwards (and downwards) from a given triangular layer
comprise all sites of the adjacent kagome layers. As $\sum^4_{i=1} \sigma_i =
0$ for each tetrahedron it follows that the magnetisations of kagome and
triangular layers are equal and opposite; iterating this, one finds that all
triangular layers have the same magnetisation. Hence, under the constraint
$\sum^4_{i=1} \sigma_i = 0$ everywhere, one can only flip strings containing at
least one spin in each layer. Such strings need not be straight: they can
meander in the kagome layers, and the entropy of this meandering in fact sets
their density at \cite{SI1}:
\begin{equation}
n_{\rm str} \sim \exp\left[ 32 h J/9 k_BT \right]~~~.
\label{eq:11}
\end{equation}

The termination of the plateau at high fields occurs by a different
mechanism. Now, the Zeeman energy $\vec{h} \cdot \hat{d}_{k(i)} \sigma_i$ wins
over the exchange, and the ground state becomes the unique saturated state with
$\sigma_1 = 1$ on the triangular layers and $\sigma_i = -1$ on the kagome
ones. A triangle with all spins $\sigma = -1$, in dimer language, does not have
a dimer attached to it. For this reason, such a configuration is called a
monomer defect -- this is now a point-like object. The energy difference
between this and a tetrahedron obeying the ice rule is given by
\begin{equation}
\Delta E = 2|J|/3 - h/3~~~.
\label{eq:hsat}
\end{equation}

At $h = h_{\rm sat} = 2J$, the (ice rule) state is degenerate with the (3-in,
1-out) state and one can choose which tetrahedron is in what state. Just as
before, this local choice gives rise to a peak in the entropy, which here is
gigantic: $S \approx 0.291 k_B$, in excess even of the Pauling entropy. 

Neglecting other (higher energy) configurations, it follows from the form of
Eq. (\ref{eq:hsat}), that the partition function, and hence all thermodynamic
quantities, near saturation depend only on the combination 
$(h-h_{\rm sat})/{T}$. Under adiabatic conditions, when $S$ is
constant,  $(h-h_{\rm sat})/{T}$ will also have a fixed
value. This is completely analogous to the case of a paramagnet, where $Z$ is a
function of $h/T$ only. 

Cooling by adiabatic demagnetisation uses this fact by reducing $h$ at fixed
$S$, so that $T$ is reduced -- in an ideal setting -- in lockstep. The same
effect occurs here, the difference being that $h$ is reduced to $h_{\rm sat}
\neq 0$ to effect cooling, and hence cooling in a finite final field becomes
possible. Indeed, the same argument applies for cooling by adiabatic {\em
  magnetisation}. As $h$ approaches $h_{\rm sat}$ {\em from below}, a drop in
$T$ also occurs! 

This theory has been put to an experimental test \cite{SIMC}, which it has
passed only partially, namely in an intermediate temperature regime. At low
$T$, the crossover captured Eq. (\ref{eq:hsat}) was found to be replaced by a
first-order transition, which terminates in a critical endpoint just like in a
liquid-gas phase diagram (Fig.~\ref{fig:liquidgas}).

This has turned out to signal a set of interesting properties of the monomer
excitations. The fundamental observation is that the Hamiltonian in
Eq. (\ref{eq:4}) is not complete -- indeed, the leading interactions between
spins are long-ranged dipolar ones \cite{SIent}. Why these have otherwise little influence
on the behaviour of spin ice is discussed in
Refs. \cite{SIPE1,SIPE2,CasMoeSon}. For our purposes, it is sufficient to note 
that two tetrahedron with (3-in, 1-out) configurations
acquire a Coulomb interaction 
\begin{equation}
v(r) \cong \frac{\mu_0}{4 \pi} \frac{\left( \vec{\mu}/a_d\right)}{r}^2~~~,
\label{eq:13}
\end{equation}
where $\vec{\mu}$ is the magnetic moment of the ions, and $a_d$ the distance
between the centers of two neighbouring tetrahedra. Note
that the prefactor contains $\mu_0$ -- this is a magnetic Coulomb interaction,
and the monomer excitations interact like magnetic charges -- or magnetic
monopoles \cite{CasMoeSon}!

Whereas the defect-free kagome ice configuration is effectively
two-dimensional, the monopoles also interact with each other between the
planes. This is thus a system where the effective dimensionality in the
presence of defects changes. The idea that low dimensionality obtains only at
low $T$ is of course commonplace. Another famous example in frustrated
magnetism is provided by Lee et al. \cite{scgodim}. 

It is even possible to generate a dimensional crossover (one-dimensional)
chains by applying a judiciously chosen magnetic field, namely one in a [110]
direction, which has zero projection on two sublattices and a finite and equal
projection onto the other two; these latter ones get pinned, and the former
point head--to--tail along chains running perpendicular to the field direction
(see Fig. \ref{fig:lattune}). The overall direction the spins of a given chain
point in is a priori undetermined, and one can thus define an effective Ising
variable $\mu_\Gamma = \pm 1$ which encodes either option for each chain
$\Gamma$. A detailed numerical study for the dipolar spin ice Hamiltonian has
found that neighbouring chains order antiferromagnetically, thus establishing
three-dimensional order at low $T$ \cite{RufMelGin}. 

We conclude this parade of field-induced effective lattice geometries with the
face-centered cubic lattice, which can be obtained if the field is rotated
away from [110] so that three sublattices are pinned by it, with the fourth
sublattice experiencing a field of a strength to cancel exactly the exchange
field due to the other three sublattices. Indeed, this is the situation one
obtains also when one tilts the field at $h_{\rm sat}$ away from [111] so that
one kagome sublattice is pinned more weakly than the other two. The
undetermined sublattice now forms a face-centred cubic lattice, and the spins on it elect
to order spontaneously in the presence of long-range interactions
\cite{RufMelGin}.   

\section{Conclusion}
\label{sec:Conc}

We hope this review provides a useful idea of the versatility that magnetic
fields offer in the study of magnetism. As always, much more has had to be
omitted than was included. Obvious topics which merit a more  detailed
treatment include magnetoelastic effects, the role of defects, properties of
more complex models systems, exotic field-tuned transitions
and of course an in-depth treatment of quantum
dynamics. Regarding the latter, we would like to mention the possibility of
generating quasiparticles with non-Abelian statistics by applying a magnetic
field \cite{Kitaev}.

\smallskip

\end{document}